# EPR INVESTIGATIONS OF DEFECTS IN $Bi_{12}GeO_{20}$:Cr SINGLE CRYSTAL IRRADIATED BY HIGH ENERGY URANIUM IONS

IRENEUSZ STEFANIUK*, PIOTR POTERA, IWONA ROGALSKA, DAGMARA WRÓBEL

Institute of Physics, University of Rzeszow, Rejtana 16a, 35-310 Rzeszow, Poland

The results of investigations of EPR spectra of chromium doped $Bi_{12}GeO_{20}$ (BGO) single crystals are presented. The crystals were studied before and after irradiation by the $^{235}U$ ions with energy 9.47 MeV/u and fluency $5\times10^{11}$ $cm^{-2}$. The effect of heating irradiated samples in air on the EPR spectra is also studied.

## INTRODUCTION

$Bi_{12}GeO_{20}$ (BGO) as well as $Bi_{12}SiO_{20}$ (BSO) crystals have been widely used in photorefractive, photoconductive, electro-optical and acousto-optical applications including two-wave mixing, four-wave mixing, phase conjugation, real-time holography, optical data storage, optical computing, electro-optical modulation, thin film optical waveguides (Rajbenbach, Huignard, 1985; Miteva, Dushkina, Gospodinov, 1995; Grachev, Kamshilin, Kobozev, Prokofiev, 2001; Kip,1998).

Earlier studies (McCullough, Harmon-Bauer, Hunt, Martin, 2001; Wardzynski, Szymczak, Pataj, Łukasiewicz, Zmija, 1982) indicate that the optical and photochromic properties of BGO:Cr crystals were due to the chromium ions located in the $Ge^{4+}$ tetrahedral positions and may occur both in $Cr^{4+}$ and $Cr^{5+}$ states. In this paper we report the EPR investigations of BGO:Cr single crystals before and after irradiation by the $^{235}U$ ions.

## EXPERIMENT

The BGO:Cr crystal was grown in the Military University of Technology by the Czochralski technique from platinum crucible. The flowing oxygen growth atmosphere was used. This BGO crystal is stoichiometric and has a lattice constant of 10.14388±0.00001 Å with a uniform composition along the axial direction. (Fu and Ozoe, 1999).

One sample (denoted # BGO-1) was irradiated by $^{235}U$ ions with energy 9.47 MeV/u (the total particles' energy was 2225 MeV) and fluency $5\times10^{11}$ $cm^{-2}$ at room temperature, without control of temperature and without cooling. Because BGO:Cr crystals exhibit strong photochromic effect, our sample was protected during irradiation by Al foil (thickness of 5 μm). This foil reduces of the particles' energy to 2120 MeV. The specifications of the samples are given in Table 1.

The sample irradiated by high energy uranium ions has cracked perpendicularly to the surface into several smaller parts. Laboratory axis system for samples # BGO-1 and BGO-2 was adopted as follows: the z-axis as parallel to the crystallographic c- axis (this reflects the way in which the sample was cut out), whereas the x- and y-axis were found experimentally based on the measurements of the EPR spectra anisotropy in the plane perpendicular to the z-axis.

The EPR spectra were measured for the non-irradiated sample a well as for the irradiated sample (before and after annealing in air) in the temperature range 140 -370 K. After measurements, the heating in air (for 20 min at temperature 500 K) was performed using a LHT 04/16 NABERTHERM furnace with C42 controller. The temperature during each heating was stable with the accuracy ±1K. The EPR spectra of the heated samples were measured after sample was cooled to the room temperature (sample # BGO-2).

---

* Corresponding author: istef@univ.rzeszow.pl

Table 1. The specifications of the samples.

| Sample # | Description |
|---|---|
| BGO-0 | BGO:Cr crystal before uranium ions irradiation |
| BGO-1 | BGO:Cr crystal after uranium ions irradiation |
| BGO-2 | Sample # BGO-1 after heating in air |

## RESULTS AND DISCUSSION

The EPR spectra of the investigated BGO:Cr samples are presented in Figs 1-3. The fine structure was clearly observed for the non-irradiated sample # BGO-0 (Fig 1). This structure is characteristic for $Cr^{4+}$ ions doped in high concentration (Wardzynski et al., 1982). The angular dependence for sample # BGO-1 at room temperature (Fig. 2) shows one strong EPR line and several weaker lines, which originate from the radiation defects created during irradiation by uranium ions. The angular dependence for sample # BGO-2 differs significantly from that for sample # BGO-1 (Fig. 3).

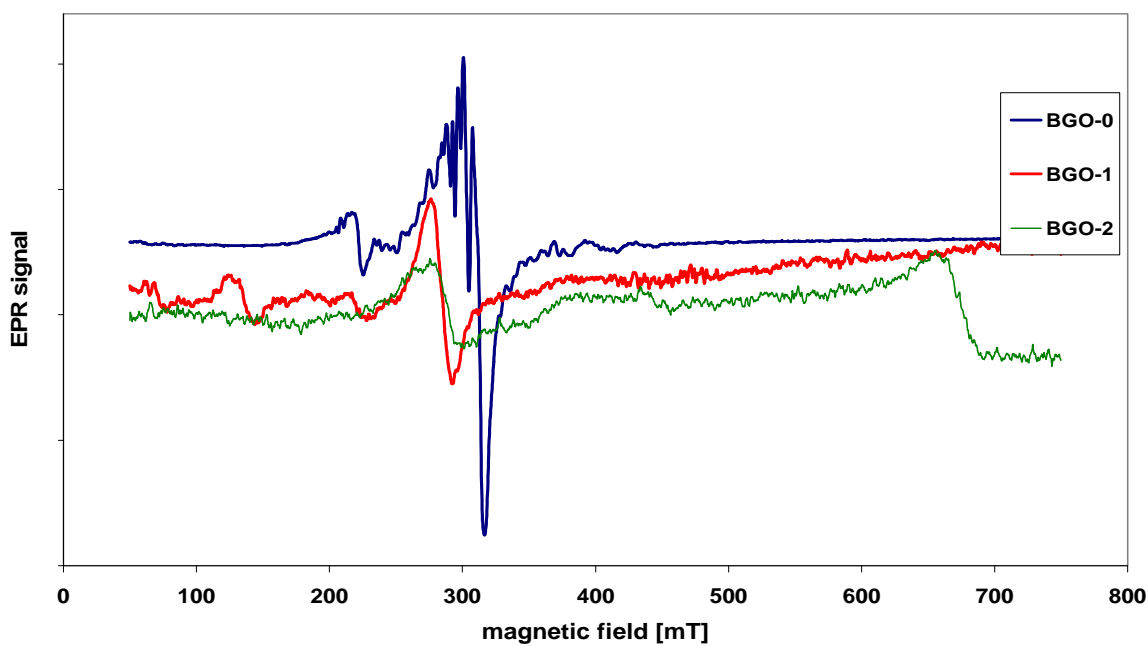

Fig. 1. The EPR spectra of sample # BGO.

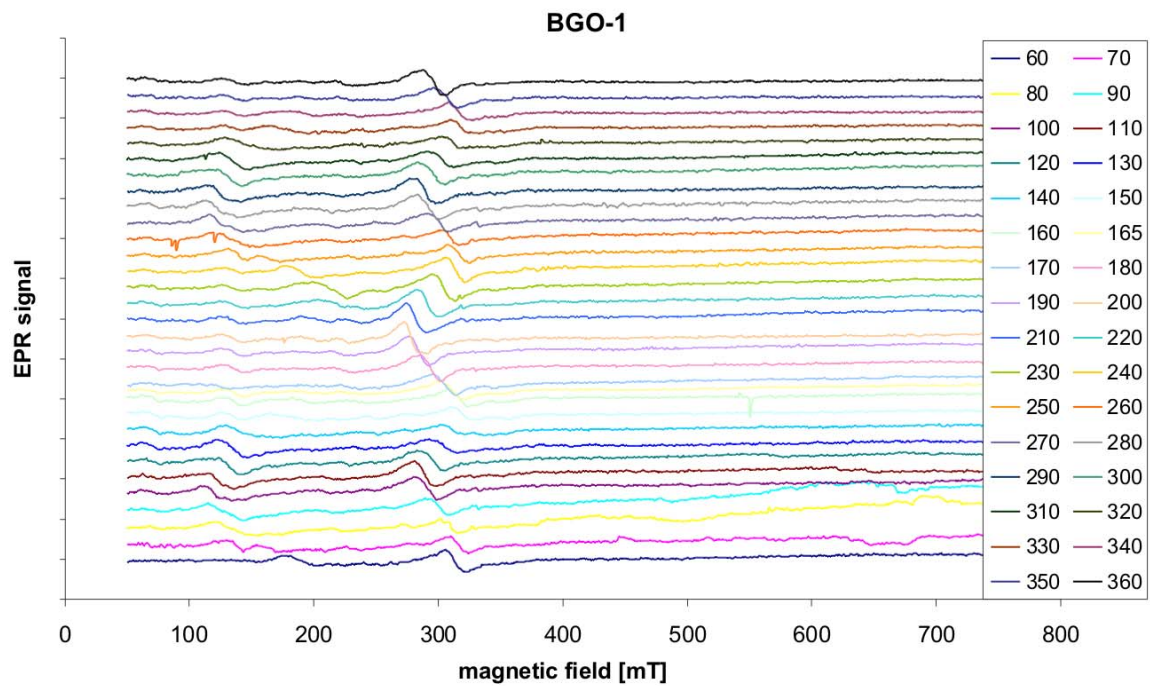

Fig. 2. The angular dependence of EPR spectra for sample # BGO-1.

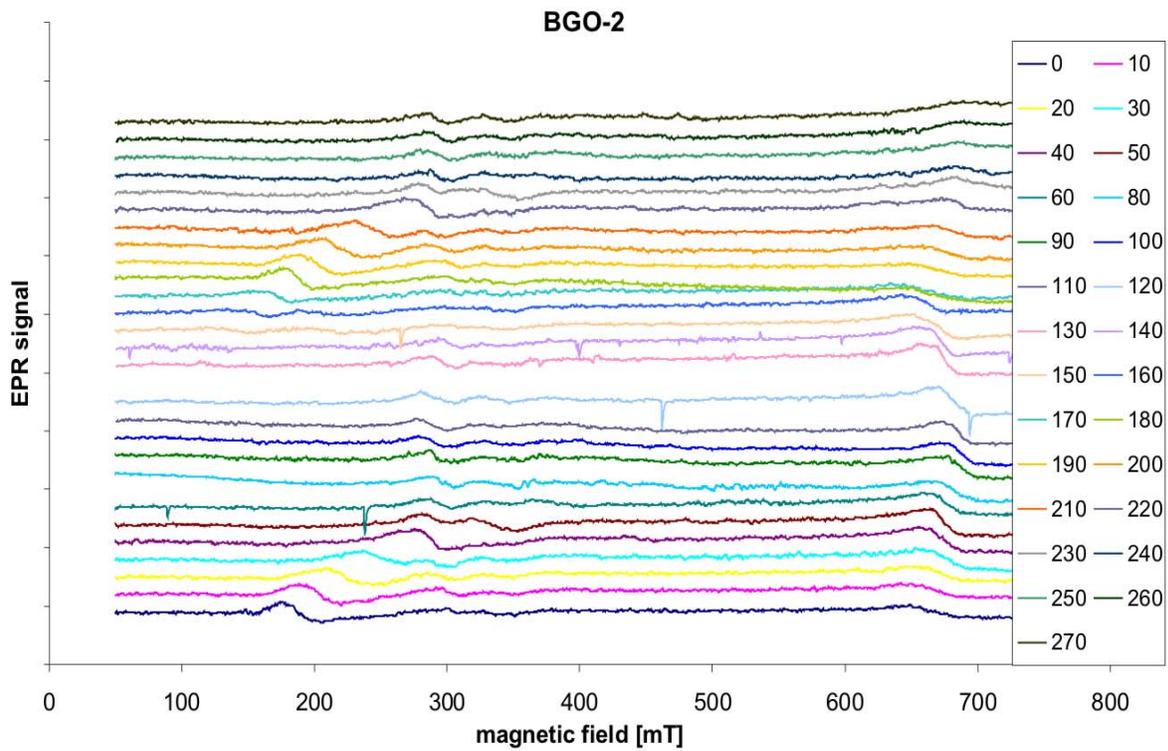

Fig. 3. The angular dependence of EPR spectra for sample # BGO-2.

Analysis of EPR spectra for sample # BGO-1 was performed using only the Zeeman part of spin Hamiltonian (Abragam, Bleaney, 1970) suitable for spin S=1/2 at arbitrary low (triclinic) symmetry. The values of the $g_{ij}$ components were obtained by fitting the experimental resonance fields to the theoretical predictions using the program EMR_NMR version 6.51 (McGavin, Mombourquette, Weil, 2002). The results are listed in Table 2.

Table 2. The fitted values of the g-tensor together with the principal values (PVs) and the orientation of the principal axes w.r.t. the laboratory axis system (*x, y, z*) for Cr ions in BGO-1 at room temperature.

| | | | | | | |
|---|---|---|---|---|---|---|
| | | | - I | | | |
| **Matrix components** | | | **PVs** | **Direction cosines** | | |
| | | | center g | | | |
| 2.159 | 0.013 | -0.079 | 2.318 | -0.433 | -0.727 | -0.534 |
| | 2.152 | -0.078 | 2.142 | -0.415 | 0.686 | -0.598 |
| | | 2.234 | 2.085 | 0.800 | -0.037 | -0.598 |
| | | | center -II | | | |
| **Matrix components** | | | **PVs** | **Direction cosines** | | |
| | | | g | | | |
| 4.954 | 0.113 | -0.137 | 5.258 | -0.470 | 0.774 | 0.424 |
| | 5.119 | -0.142 | 4.911 | -0.787 | -0.585 | 0.196 |
| | | 4.818 | 4.721 | 0.399 | -0.242 | 0.884 |

For samples # BGO-1 and BGO-2 the temperature dependence of the $g_{eff}$ factor was determined. The changes in $g_{eff}$ were observed after treatment, namely, for sample # BGO-2 annealed after irradiation the change in the temperature dependence was stronger than for sample # BGO-1 irradiated only. The analysis of the fitted g-tensor principal values (PVs) and the orientation of the principal axis system (PAS) for the two Cr centers in BGO-1 reveals their different orientation w.r.t. the laboratory axis system (*x, y, z*). This indicates that presumably different kinds of radiation defects are created in BGO crystals during irradiation. After heating of sample # BGO-1 in air (500 K, 20 min) two low-field EPR lines observed before heating near 120 and 220 mT disappeared, whereas only one EPR line near 300 mT remained visible for the center -I.

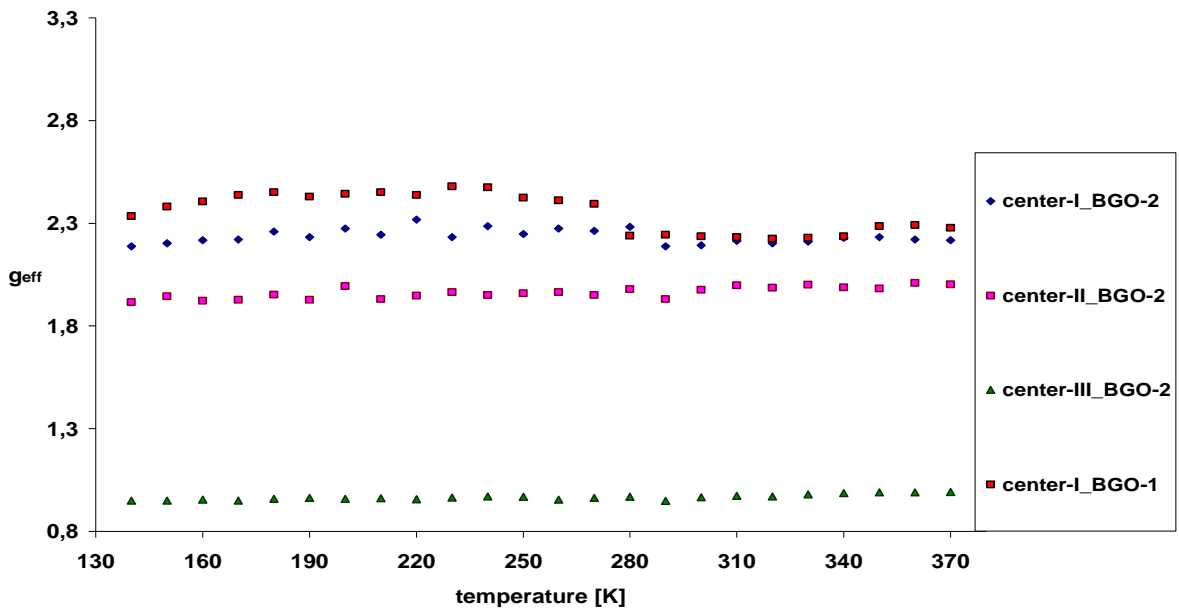

Fig. 4. The temperature dependence of the $g_{eff}$ values for sample # BGO-1 and BGO-2.

Aditionally, after treatment, two new EPR lines appear - a weak line near 340 mT and a strong line near 670 mT. The observed changes are probably due to the change in the valence of chromium ions. To confirm this assertion, further detailed study is necessary. Analysis of the temperature dependence of the obtained $g_{eff}$ values for both samples (Fig. 4) indicates a slight decrease in $g_{eff}$ with decreasing temperature. Only for the center - I for sample # BGO-1, a slight increase in $g_{eff}$ near 280 K was observed. The variation of the $g_{eff}$ value with temperature is small as compared with that for the angular dependence of EPR spectra, especially for sample # BGO-1.

## CONCLUSIONS

The g-tensor value for the two paramagnetic centers (center-I and center-II) in sample # BGO-1 was determined. The results indicate that irradiation creates two different types of defect centers. One center is thermally instable, but another one is resistant to heating.

After BGO:Cr irradiation, the fine structure of EPR spectra of chromium ions disappear. Irradiation of BGO:Cr crystals by uranium ions leads to the change of the structure of EPR spectra. The annealing of the irradiated sample leads to the changes of the line positions and creation of a new line in higher magnetic field.